\documentclass[12pt]{article}
\usepackage{amsfonts, amsmath, amsthm, amssymb,graphicx,natbib}
\title{\textbf{An algorithm for the multivariate group lasso with covariance estimation}}
\author{Ines Wilms\footnote{Corresponding author: ines.wilms@kuleuven.be} \ and Christophe Croux \\
\footnotesize{\textit{Leuven Statistics Research Centre, KU Leuven, Belgium}} }
\date{ }
\usepackage[left=1in,top=1in,right=1in,bottom=1in,nohead,paperwidth=8.5in, paperheight=11in]{geometry} %1 inch margins, 8 1/2 x 11-inch pages
\newtheorem{mydef}{Lemma}

\begin{document}
\maketitle

\noindent
\textbf{Abstract}
We study a group lasso estimator for the multivariate linear regression model that accounts for correlated error terms. 
A block coordinate descent algorithm is used to compute this estimator. 
We perform a simulation study with categorical data and multivariate time series data, typical settings with a natural grouping among the predictor variables.
Our simulation studies show the good performance of the proposed group lasso estimator compared to alternative estimators. We illustrate the  method on a time series data set  of gene expressions.
\bigskip


\newpage
\section{Introduction}
% Introduction grouplasso
Since its introduction by \cite{Yuan06}, the group least absolute shrinkage and selection operator (group lasso) has received considerable interest in the statistical literature (e.g. \citealp{Meier07}, \citealp{Wang08}, \citealp{Simon13}, \citealp{Alfons15}). 
In many applications, the parameter vector in the regression model is structured into groups. Typical examples are (i) regression with categorical variables, where a group of dummies represents each categorical variable, or (ii) time series regression where several lagged values of the same time series are included in the model. In settings with such a natural group structure, one wants to select either all or none of the variables belonging to a particular group. The key strength of the group lasso lies in its ability to perform such groupwise selection.

% Multivariate regression model : need for regularization
We consider the group lasso for the multivariate linear regression model. The multivariate linear regression model generalizes the classical linear regression model in that it regresses $q>1$ responses instead of a single response on $p$ predictors. Let ${\bf Y}=({\bf y}_1,\ldots,{\bf y}_q) \in \mathbb{R}^{n \times q}$ be the response matrix, and ${\bf X}=({\bf x}_1,\ldots,{\bf x}_p) \in \mathbb{R}^{n \times p}$ be the predictor matrix. The error vectors are assumed to follow a normal $N_q(0,\boldsymbol \Sigma)$ distribution, with $\boldsymbol \Sigma^{-1}=\boldsymbol \Omega$, and are collected in the columns of  the error matrix ${\bf E}$.  The multivariate linear regression model is given by
\begin{equation}
{\bf Y} = {\bf X} {\bf B} + {\bf E}, \label{model}
\end{equation}
where ${\bf B} \in  \mathbb{R}^{p \times q}$ is the coefficient matrix. We assume that this coefficient matrix contains $K$ predefined groups. Denote each group as ${\bf B}_{G_j}$ where $j \in \{1,\ldots,K\}$. 

Recently, \cite{Li15} discussed the group lasso for the multivariate linear regression model. Their multivariate group lasso estimator\footnote{Note that \cite{Li15} consider a more general version of the multivariate group lasso that also allows for selection of  predictors within the important groups.} is given by
\begin{equation}
\widehat{\bf{B}} = \underset{\bf{B}}{\operatorname{argmin }} \ \dfrac{1}{2n} \operatorname{tr} \left(  ({\bf Y -{\bf X} {\bf B}} )^T ({\bf Y -{\bf X} {\bf B}} ) \right) +   \sum_{j=1}^{K} \lambda_{G_j}m_j ||{\bf B}_{G_j} ||_2, \label{penloglikLi}
\end{equation}
where $\operatorname{tr}(\cdot)$ denotes the trace, $\lambda_{G_j}>0,$  for $1\leq j\leq K,$ are sparsity parameters, and $m_j$ equals the number of elements in group $j$. 
A groupwise penalty is used for the regression coefficients. As such, variables are selected in a grouped manner: either all elements of a certain group are set to zero or none.
%Furthermore, the adaptive group $L_1$ penalty is consistent for variable selection, whereas the group $L_1$ penalty is not \citep{Wang08}.

However, \cite{Li15} do not account for correlated errors. We extend the multivariate group lasso from \cite{Li15}
such that the correlation between the error terms of the different equations of the multivariate regression model is taken into account. To this end, we simultaneously estimate the regression parameters $\bf{B}$ and the inverse covariance matrix of the error terms $\boldsymbol{\Omega}$ using penalized maximum likelihood: %The minimizer of the penalized negative log-likelihood for the multivariate linear regression model is given by
\begin{equation}
(\widehat{\bf{B}}, \widehat{\boldsymbol{\Omega}}) = \underset{(\bf{B}, \boldsymbol{\Omega})}{\operatorname{argmin }} \ \dfrac{1}{2n} \operatorname{tr} \left(  ({\bf Y -{\bf X} {\bf B}} )^T ({\bf Y -{\bf X} {\bf B}} ) \boldsymbol \Omega \right)  - \dfrac{1}{2}\operatorname{log}|\boldsymbol{\Omega}|+   \sum_{j=1}^{K} \lambda_{G_j}m_j ||{\bf B}_{G_j} ||_2 + \lambda_{\omega} \sum_{k \neq k^{\prime}} |\omega_{kk^{\prime}}|, \label{penloglik}
\end{equation}
where $\lambda_\omega>0$ is a sparsity parameter, and $\omega_{kk}$ is the $k^{th}$ element of $\boldsymbol{\Omega}$. We use an $L_1$ penalty for the elements of the inverse covariance matrix.

Section \ref{Algorithm} describes the algorithm used to approximate the minimizer of the objective function in \eqref{penloglik}. The main modification in the algorithm compared to the proposal of \cite{Li15} is that the error covariance structure is taken into account. 
Simulation studies are performed in Section  \ref{Simulation}. Our simulations show that the group lasso with covariance estimation considerably outperforms the group lasso without covariance estimation. 
Section \ref{Application} contains a real data example. 
%Finally, Section \ref{Conclusion} concludes.

\section{The algorithm \label{Algorithm}}
To find the minimum of the penalized negative log-likelihood in \eqref{penloglik}, we iteratively solve for ${\bf B}$ conditional on $\boldsymbol \Omega$ and for $\boldsymbol \Omega$ conditional on ${\bf B}$.

\bigskip

\noindent
\textit{Solving for ${\bf B}$ conditional on $\boldsymbol \Omega$.}
When $\boldsymbol \Omega$ is fixed, the minimization problem in \eqref{penloglik} is equivalent to 
\begin{equation}
\widehat{\bf{B}} = \underset{\bf{B}}{\operatorname{argmin }} \ \dfrac{1}{2n} \operatorname{tr} \left(  ({\bf Y -{\bf X} {\bf B}} )^T ({\bf Y -{\bf X} {\bf B}} ) \boldsymbol \Omega \right) + \sum_{j=1}^{K} \lambda_{G_j} m_j ||{\bf B}_{G_j} ||_2. \label{solveB}
\end{equation}

\noindent
To find a solution to \eqref{solveB}, we use a block coordinate descent algorithm, analogously to \cite{Friedman07} for solving the single response lasso problem, or to \cite{Li15} for the multivariate group lasso problem without covariance estimation.
Lemma 1 (Lemma 4.2 from Chapter 4 in \citealp{Buhlmann11}) provides a necessary and sufficient condition for $\widehat{\bf{B}}$  to be a solution of \eqref{solveB}.

\begin{mydef} \label{Lemma1}
Denote the loss function by 
\begin{equation}
\rho({\bf B})= \dfrac{1}{2n} \operatorname{tr} \left(  ({\bf Y -{\bf X} {\bf B}} )^T ({\bf Y -{\bf X} {\bf B}} ) \boldsymbol \Omega \right). \nonumber
\end{equation}
The gradient of the loss function evaluated at $\bf{B}$ is 
\begin{equation} \label{gradient}
\nabla \boldsymbol\rho(\bf{B}) = - \dfrac{1}{n} {\bf X}^T ({\bf Y} - {\bf X}\bf{B})\boldsymbol{\Omega}. \nonumber
\end{equation} 
A necessary and sufficient condition for $\bf{B}$  to be a solution of \eqref{solveB} is 
\begin{enumerate}
\item $\nabla \boldsymbol\rho({\bf{B})}_{G_j} + \lambda_{G_j}m_j \dfrac{{\bf{B}}_{G_j}}{||{\bf{B}}_{G_j}||_2} = \bf{0}$  if ${\bf{B}}_{G_i} \neq {\bf 0}$
\item $||\nabla \boldsymbol\rho({\bf{B})}_{G_j}||_2 \leq \lambda_{G_j}m_j$ if  ${\bf{B}}_{G_j} = {\bf 0}$.
\end{enumerate} 
\end{mydef}

To start up the block coordinate descent algorithm, an initial value for ${\bf B}$ is needed. We use the lasso estimator obtained by performing $q$ separate lasso regressions. 
Assume now that $\widehat{{\bf{B}}}^{(m-1)}$ is given, for $m\geq 1$. 
In the following iteration step $m$, we update  our estimate from $\widehat{{\bf{B}}}^{(m-1)}$ to $\widehat{{\bf{B}}}^{(m)}$.
Note that the $ik^{th}$ element of the gradient of the loss function evaluated at ${\bf{B}}$ is given by
\begin{eqnarray}
\nabla \boldsymbol\rho({\bf{B})}_{ik} & = &  - \dfrac{1}{n} {\bf x}_i^T ({\bf Y - {\bf X}{\bf B}})\boldsymbol\Omega_k  \nonumber\\
 & = & \dfrac{1}{n} \left( -{\bf x}_i^T({\bf Y - {\bf X}{\bf B}}^{-ik})\boldsymbol\Omega_k  + \omega_{kk} ||{\bf x}_i||^2_2 B_{ik}\right) \nonumber \\
 & =& \dfrac{1}{n} \left(-{\bf{S}}_{ik} + \omega_{kk} ||{\bf x}_i||^2_2 B_{ik} \right), \nonumber
\end{eqnarray}
with ${\bf x}_i$ the $i^{th}$ column of ${\bf X}$, $\boldsymbol \Omega_k$  the $k^{th}$ row of $\boldsymbol{\Omega}$, $\omega_{kk}$  the $kk^{th}$ element of $\boldsymbol{\Omega}$, ${\bf B}^{-ik}$ is $\bf B$ with element $ik$ replaced by zero, and ${\bf{S}}_{ik}={\bf x}_i^T({\bf Y - {\bf X}{\bf B}}^{-ik})\boldsymbol\Omega_k$.

In iteration step $m$, we cycle through all groups $G_j,$ with  $j=1,\ldots,K$. 
If, for group $G_j$ $$||\nabla \boldsymbol\rho(\widehat{\bf{B}}^{(m-1)})_{G_j}||_2 \leq \lambda_{G_j}m_j$$ holds, then according to condition 2 from Lemma \ref{Lemma1}, all elements of group $G_j$ of $\widehat{{\bf{B}}}^{(m)}$ are set to zero. Otherwise, according to condition 1 from Lemma \ref{Lemma1}, for every element $ik$ of ${\bf B}$ belonging to group $G_j$ it needs to hold that 
\begin{eqnarray}
 \ \ \ 0 & = & \nabla \boldsymbol\rho({\bf{B}})_{ik} + \lambda_{G_j}m_j \dfrac{B_{ik} }{||{\bf{B}}_{G_j}||_2}  \nonumber\\
\iff \ \  \ 0 & = & \dfrac{-{\bf{S}}_{ik}}{n} + \dfrac{\omega_{kk} ||{\bf x}_i||^2_2 }{n}B_{ik} +  \dfrac{\lambda_{G_j}m_j }{||{\bf{B}}_{G_j}||_2}{B}_{ik}   \nonumber \\
\iff {B}_{ik} &=& \dfrac{{\bf{S}}_{ik}}{\omega_{kk} ||{\bf x}_i||^2_2 + \dfrac{n\lambda_{G_j}m_j}{||{\bf{B}}_{G_j}||_2}}.  \label{Bupdate}
\end{eqnarray}
%where we use from equation \eqref{gradient} that
%\begin{eqnarray}
%\nabla \boldsymbol\rho(\widehat{\bf{B}})_{jk} & = &  - \dfrac{1}{n} {\bf x}_j^T ({\bf Y - {\bf X}{\bf \widehat{B}}})\boldsymbol\Omega_k  \nonumber\\
% & = & \dfrac{1}{n} \left( -{\bf x}_j^T({\bf Y - {\bf X}{\bf \widehat{B}}}^{-jk})\boldsymbol\Omega_k  + \omega_{kk} ||{\bf x}_j||^2_2 \widehat{B}_{jk}\right) \nonumber \\
% & =& \dfrac{1}{n} \left(-\widehat{\bf{S}}_{jk} + \omega_{kk} ||{\bf x}_j||^2_2 \widehat{B}_{jk} \right), \nonumber
%\end{eqnarray}
%with ${\bf x}_j$ the $j^{th}$ column of ${\bf X}$, $\boldsymbol \Omega_k$  the $k^{th}$ row of $\boldsymbol{\Omega}$, $\omega_{kk}$  the $kk^{th}$ element of $\boldsymbol{\Omega}$, ${\bf \widehat{B}}^{-jk}$ is $\widehat{\bf B}$ with element $jk$ replaced by zero, and $\widehat{\bf{S}}_{jk}={\bf x}_j^T({\bf Y - {\bf X}{\bf \widehat{B}}}^{-jk})\boldsymbol\Omega_k$.
%\smallskip

Note that the right-hand-side from equation \eqref{Bupdate} involves $B_{ik}$ in the computation of $||{\bf{B}}_{G_j}||_2$. For this, we use the estimate from the previous iteration. %$\widehat{\bf{B}}_{G_i}^{(m-1)}$ to update $\widehat{B}_{jk}$ from $\widehat{B}_{jk}^{(m-1)}$ to $\widehat{B}_{jk}^{(m)}$ according to equation \eqref{Bupdate}. %, thereby following \cite{Li15}. %An alternative is to iteratively update equation \eqref{Bupdate}, which can be time consuming.
Table \ref{algorithm} provides a schematic overview of the block coordinate descent algorithm.
\smallskip

\begin{table}
\caption{Block Coordinate Descent Algorithm to solve for ${\bf B}$ conditional on $\boldsymbol \Omega$. \label{algorithm}}
\small
\begin{tabular}{lll} \hline
1: & \multicolumn{2}{l}{\textbf{Initialization} Let ${\bf B}^{(0)}$ be an initial parameter estimate. We use the lasso estimator } \\
 & \multicolumn{2}{l}{obtained by performing $q$ separate lasso regressions. Set $m=0$. } \\
2: & \multicolumn{2}{l}{\textbf{Repeat}} \\
 & \multicolumn{2}{l}{$m \leftarrow m+1$}\\
 & \multicolumn{2}{l}{For each block $j=1,\ldots,K$:} \\
&& If $||\nabla \boldsymbol\rho(\widehat{\bf{B}}^{(m-1)})_{G_j}||_2 \leq \lambda_{G_j}m_j$: set $\widehat{\bf{B}}^{(m)}_{G_j} = {\bf 0}$ \\
 && Else: Update every $ik^{th}$ element $\widehat{B}^{(m)}_{ik}$ of $\widehat{\bf{B}}^{(m)}$ belonging to group $G_j$ by  \\
 && \hspace{2cm} $\widehat{B}^{(m)}_{ik} = \dfrac{\widehat{S}^{(m-1)}_{ik}}{\omega_{kk} ||{\bf x}_i||_2^2 + \dfrac{n\lambda_{G_j}m_j}{||\widehat{\bf{B}}_{G_j}^{(m-1)}||_2} }$.\\
% && \hspace{2cm} where $\widehat{S}_{jk}={\bf x}_j^T ({\bf Y} - {\bf X} \widehat{{\bf B}}^{(m-1), -jk})\boldsymbol \Omega_k$, with ${\bf x}_j$ the $j^{th}$ column of ${\bf X}$, $\boldsymbol \Omega_k$ the  \\
% &&  \hspace{2cm} $k^{th}$ row of $\boldsymbol{\Omega}$, and $\widehat{{\bf B}}^{(m-1),-jk}$ is $\widehat{\bf B}^{(m-1)}$ with element $jk$ replaced by zero.\\
3: & \multicolumn{2}{l}{\textbf{Until} convergence. We iterate intil the relative change in the value of the objective function} \\
 & \multicolumn{2}{l}{ in \eqref{solveB} in two successive iterations is smaller than the tolerance value $\epsilon=10^{-2}.$} \\ \hline
\end{tabular}
\end{table}

Note that the estimator in \eqref{solveB} is a multivariate \textit{adaptive} group lasso estimator since each group has its own sparsity parameter $\lambda_{G_j}$.  We take $\lambda_{G_j}=\lambda/||\widehat{{\bf B}}^{(0)}_{G_j} ||_2$, for $j=1,\ldots,K$. This way, only one tuning parameter for the regression coefficients needs to be selected instead of $K$. %The optimal values of the sparsity parameter $\lambda$ can be selected by different criteria. 
We use a grid of sparsity parameters and search for the optimal one using the Bayesian Information Criterion (BIC). %, see \cite{Abegaz13} and references therein.  
The BIC is given by
\begin{equation}
BIC_{\lambda} = -2 \operatorname{log L}_{\lambda} + k_{\lambda} \operatorname{log}(n), \nonumber
\end{equation}
where $\operatorname{log  L}_{\lambda}$ is the estimated log-likelihood, corresponding to the first term of the objective function in \eqref{solveB}, using sparsity parameter $\lambda$, and $k_\lambda$ is the number of non-zero estimated regression coefficients.

\bigskip

\noindent
\textit{Solving for $\boldsymbol \Omega$ conditional on ${\bf B}$.}
When ${\bf B}$ is fixed, the minimization problem
in \eqref{penloglik} corresponds to the graphical lasso \citep{Friedman08} on the residuals ${\bf Y} - {\bf X} {\bf B}$. We use the Bayesian Information Criterion to select the optimal value of the sparsity parameter $\lambda_{\omega}$ (e.g. \citealp{Yuan07}).

\bigskip

\noindent
\textit{Starting value and convergence.} We start by taking  $\boldsymbol \Omega= \bf I$ and then iteratively solve for ${\bf B}$ conditional on $\boldsymbol \Omega$ and for $\boldsymbol \Omega$ conditional on ${\bf B}$. We iterate until the relative change in the value of the objective function in \eqref{penloglik} in two successive iterations is smaller than the tolerance value $\epsilon=10^{-2}.$

\section{Simulation \label{Simulation}}
We compare the performance of the multivariate group lasso with covariance estimation, ``\verb|GroupLasso+Cov|", to 
\begin{enumerate}
\item The multivariate group lasso without covariance estimation, ``\verb|GroupLasso|", i.e. the solution of \eqref{penloglikLi}, 
\item The multivariate lasso with covariance estimation, ``\verb|Lasso+Cov|", i.e. the solution of \eqref{penloglik} with $m_j=1,$ for $ 1 \leq j \leq K$, where $K=p\times q$. The resulting estimator is equivalent to the Multivariate Lasso With Covariance Estimator introduced in \cite{Rothman10}.
\item The multivariate lasso without covariance estimation, ``\verb|Lasso|", i.e. the solution of \eqref{penloglikLi} with $m_j=1,$ for $ 1 \leq j \leq K,$ where $K=p\times q$.
\end{enumerate}
Note that ``\verb|Lasso+Cov|" and ``\verb|Lasso|" do not take the group structure among the predictors into account.

\subsection{Predictor groups}
The first data configuration corresponds to a regression model with categorical predictors,
the second to a time series model. 
\bigskip

\noindent
{\bf Categorical data.}
We consider a design  similar to model I from \cite{Yuan06} for the univariate regression model. 
We generate a sample $Z_{ij}$, for $i=1,\ldots,n$ and $j=1,\ldots,K$, of size $n$ from a centered multivariate normal distribution with covariance matrix $\boldsymbol{\Sigma}^Z$ where   
%We first generate $K=\{5,20,50\}$ latent variables $Z^1,\ldots,Z^K$ each of length $n=50$ according to a centered multivariate normal distribution with covariance matrix $\boldsymbol{\Sigma}^Z$ where 
\begin{equation}
\Sigma^Z_{ij}=0.5^{|i-j|}. \nonumber
\end{equation}
Afterwards, $Z_{ij}$ is trichotomized as 
\begin{equation}
C_{ij} = \begin{cases}
0  & \text{if} \hspace{0.5cm} Z_{ij} < \Phi^{-1}(\dfrac{1}{3}) \\
1  & \text{if} \hspace{0.5cm} Z_{ij} > \Phi^{-1}(\dfrac{2}{3}) \\
2  & \text{if} \hspace{0.5cm} \Phi^{-1}(\dfrac{1}{3}) < Z_{ij} < \Phi^{-1}(\dfrac{2}{3}),
\end{cases} \nonumber
\end{equation}
for $i=1,\ldots,n=50$ and $j=1,\ldots,K$, where $K$ denotes the number of groups. We take $K \in\{5, 20, 50\}$.
The $(n\times p)$ matrix of predictors ${\bf X}$ then contains in its columns the $p=2K$ dummy variables $D^{0}_{ij}= \operatorname{I}(C_{ij}=0)$ and  $D^{1}_{ij}= \operatorname{I}(C_{ij}=1)$, for $j=1,\ldots,K$ and $i=1,\ldots,n$, where $\operatorname{I}(\cdot)$ is the indicator function.
Next, the $q=5$ responses are simulated from
\begin{equation}
{\bf Y} = {\bf B} {\bf X} + {\bf E},
\end{equation}
where ${\bf B}={\bf I}_q \otimes {\bf b}$, with ${\bf b}= (2,-1,\ldots,2,-1)$ a vector of length $p/q$. For the error covariance matrix $\boldsymbol \Sigma$ we consider different structures, detailed in the Section \ref{Sigma}. 
The group lasso accounts for the grouped predictor variables by selecting either all or none of the dummy variables corresponding to a particular categorical variable in one of the equations of the multivariate regression model.

\bigskip

\noindent
{\bf Time series.}
We generate the data from a VAR(2) model 
\begin{equation}
{\bf y}_t = {\bf B}_1 {\bf y}_{t-1} + {\bf B}_2 {\bf y}_{t-2} + {\bf e}_t,
\end{equation}
for $t=1,\ldots,T=50$, where ${\bf y}_t$ is a $q$-dimensional vector, with $q \in\{5,20,50\}$. The coefficient matrices ${\bf B}_1$ and ${\bf B}_2$ have the same sparse structure and ${\bf e}_t \sim N_q({\bf 0},\boldsymbol \Sigma)$. For the error covariance matrix $\boldsymbol \Sigma$ we consider different structures, detailed in Section \ref{Sigma}. 

The above model is a Vector AutoRegressive (VAR) model of order two since two lagged values of each time series are included as predictors. The group lasso accounts for the grouped predictor variables by selecting either all or none of the lagged values of a particular time series in one of the equations of the VAR. As a result, ${\bf \widehat{B}}_1$ and ${\bf \widehat{B}}_2$  have their zero elements in exactly the same cells.

We generate the sparse coefficient matrices ${\bf B}_1$ and ${\bf B}_2$ from a network structure (see \citealp{Fujita07}). %The sparsity structure of the coefficient matrices ${\bf B}_1$ and ${\bf B}_2$  is derived from the adjacency matrix ${\bf A}$, see for instance \cite{Fujita07}. 
This dimensions of this network are similar to the ones in the real data example to be discussed in Section \ref{Application}.
The adjacency matrix ${\bf A}$ represents the network structure where the nodes are the $q$ different time series. Element $A_{ij}=1$ if a  directed edge is drawn from node $i$ to node $j$, otherwise $A_{ij}=0$. 
To construct the adjacency matrix  ${\bf A}$, we start (iteration $l=0$) from a network of two randomly selected nodes that are connected with a bidirectional edge. Next, in iteration $l=1,\ldots,q-2$, a node that is currently not in the network is randomly selected. This new node is connected to a node that is present in the network via an edge whose direction is randomly chosen. The probability 
$$
\pi^{(l-1)}_m = \dfrac{d^{(l-1)}_m}{\sum_n d^{(l-1)}_n},
$$
that the new node is connected to node $m$ depends on the  degree $d^{(l-1)}_m$ of the node present in the network from iteration $l-1$.  The degree of a node equals the number of edges starting from it. %The probability $\pi_i$ is given by
%$$
%\pi_i = \dfrac{d_i}{\sum_j d_j}.
%$$
Finally, we set ${\bf B}_1=0.4{\bf A}$ and ${\bf B}_2=0.2{\bf A}$.

\subsection{Structure of the error terms \label{Sigma}}
We consider three structures for the error covariance matrix $\boldsymbol\Sigma$ and its inverse $\boldsymbol\Omega$, see e.g. \cite{Rothman10}:
\begin{enumerate}
\item \textbf{Sparse} $\boldsymbol \Omega$: $\Sigma_{ij}= \rho^{|i-j|}$, with $\rho \in\{0.2,0.4,0.6,0.8\}$. The error covariance matrix $\boldsymbol \Sigma$ is a dense matrix, whereas its inverse $\boldsymbol{\Omega}$ is a band matrix.
\item \textbf{Diagonal} $\boldsymbol \Omega$: $\boldsymbol\Sigma={\bf I}_q$. Both the error covariance matrix and its inverse are diagonal.
\item \textbf{Dense} $\boldsymbol \Omega$: $\Sigma_{ij}= 0.5( (|i-j|+1)^{2\times0.9} - 2 |i-j|^{2\times0.9} + (|i-j| - 1)^{2\times0.9} )$. Both the error covariance matrix and its inverse have a dense structure.
\end{enumerate}

\subsection{Performance measures}
We measure estimation accuracy by looking at the Mean Absolute Estimation Error given by
\begin{equation}
\text{MAEE} = \dfrac{1}{N} \dfrac{1}{p \times q}  \sum_{m=1}^{N}\sum_{j=1}^{q}\sum_{i=1}^{p}  | \widehat{b}^{(m)}_{ij} - b_{ij} |,
\end{equation}
where $\widehat{b}^{(m)}_{ij}$ is the estimate of the $ij^{th}$ element of ${\bf B}$ in simulation run $m$. We take $N=1000$ simulation runs.
\smallskip

\noindent
We measure sparsity recognition  by looking at the True Positive Rate and the True Negative Rate given by
\begin{gather}
\text{TPR} = \dfrac{1}{N}  \sum_{m=1}^{N}\dfrac{  \# \{ (i,j) :  \hat{b}^{(m)}_{ij} \neq 0 \  and \ b_{ij} \neq 0 \}}{\# \{ (i,j,) :  \ b_{ij} \neq 0 \}} \nonumber \\
\text{TNR} = \dfrac{1}{N}  \sum_{m=1}^{N}\dfrac{  \# \{ (i,j) :  \hat{b}^{(m)}_{ij} = 0 \  and \ b_{ij} = 0 \}}{\# \{ (i,j,) :  \ b_{ij} = 0 \}}. \nonumber
\label{sparsityperformance} 
\end{gather}
TPR gives the hit rate of including an important variable, whereas the TNR gives the hit rate of excluding an unimportant variable. Both should be as large as possible for reliable variable selection.

\subsection{Results}
In this section, we discuss the results for the two data configurations. We show that the \verb|GroupLasso+Cov| considerably improves the \verb|GroupLasso| as soon as the errors are correlated.

\bigskip

\begin{figure}[t]
\centering
\includegraphics[width=12cm]{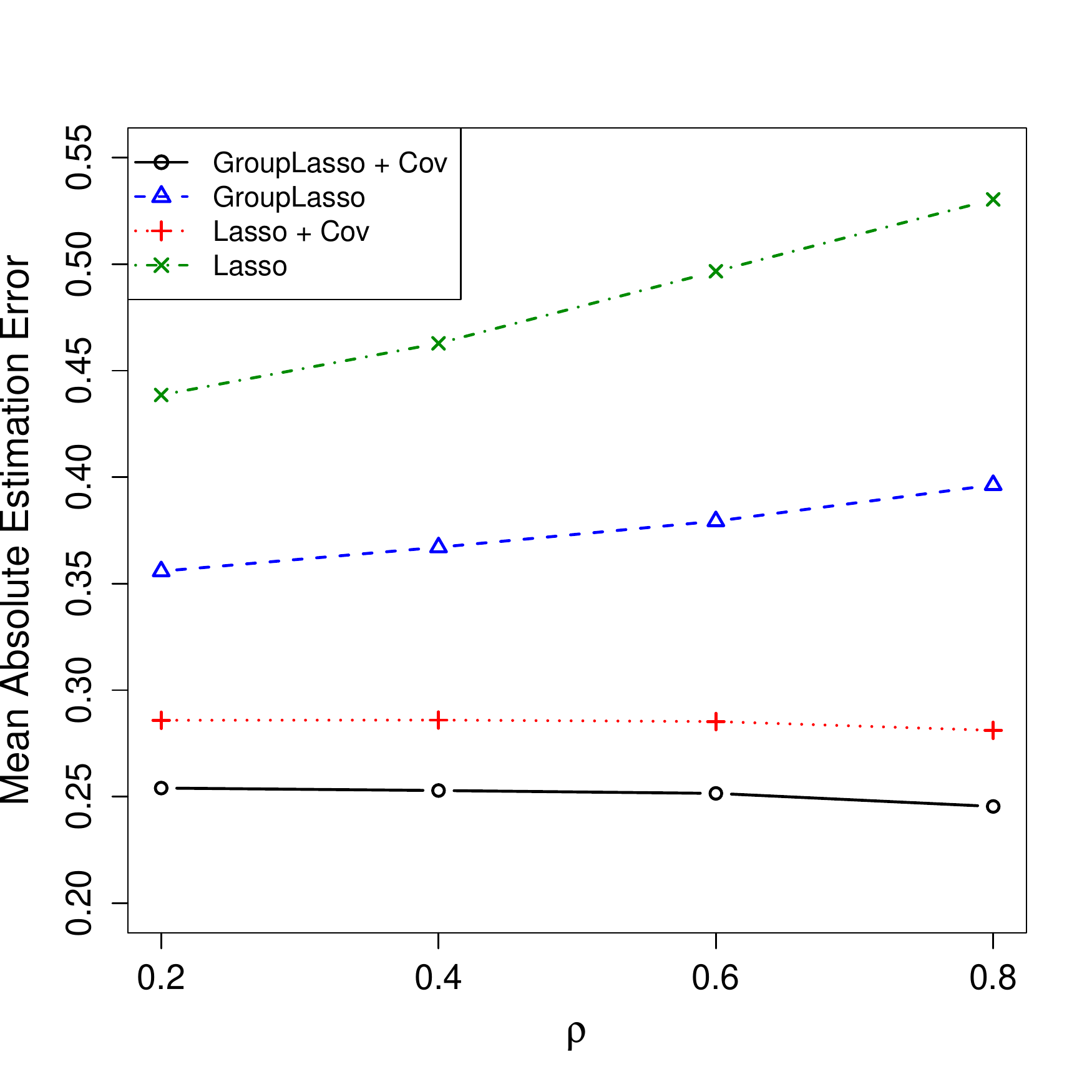}
\caption{\small Multivariate regression with $q=5$ responses, $K=5$ categorical regressors and $n=50$: Mean Absolute Estimation Error versus the correlation $\rho$, for the four considered estimators. \label{MAEECATrho}}
\end{figure}

\noindent
{\bf Categorical data.}
We first discuss the results for the sparse inverse error covariance structure (cfr. Section \ref{Sigma}). The MAEE with $K=5$ categorical regressors is displayed in Figure \ref{MAEECATrho} for different values of the correlation $\rho$. Similar conclusion can be made for $K=20$ or $K=50$ categorical regressors and are, hence, omitted.

The \verb|GroupLasso+Cov| substantially outperforms the \verb|GroupLasso| for all values of the correlation $\rho$. The margin by which the former outperforms the latter increases when $\rho$ increases.  The \verb|GroupLasso+Cov| achieves this improved estimation accuracy since it accounts for the error correlation whereas the  \verb|GroupLasso| does not. 
%A similar trend occurs when comparing the \verb|Lasso+Cov| to the \verb|Lasso|.
Besides, as expected for grouped predictors, the group lasso estimators outperform the corresponding lasso estimators. 
%When the predictors variables are grouped, the distinct feature of the group lasso that accounts for this group structure increases estimation accuracy.

\begin{table}[t]
\small
\caption{\small Multivariate regression with $q=5$ responses, $K \in \{5, 20, 50\}$ categorical regressors and $n=50$: Mean Absolute Estimation Error, True Positive and True Negative Rate. \label{CatTable}}

\resizebox{0.88\textwidth}{!}{\begin{minipage}{\textwidth}
\centering
\begin{tabular}{lllcccccc} \hline
 &&& \multicolumn{2}{c}{\includegraphics[width=4cm]{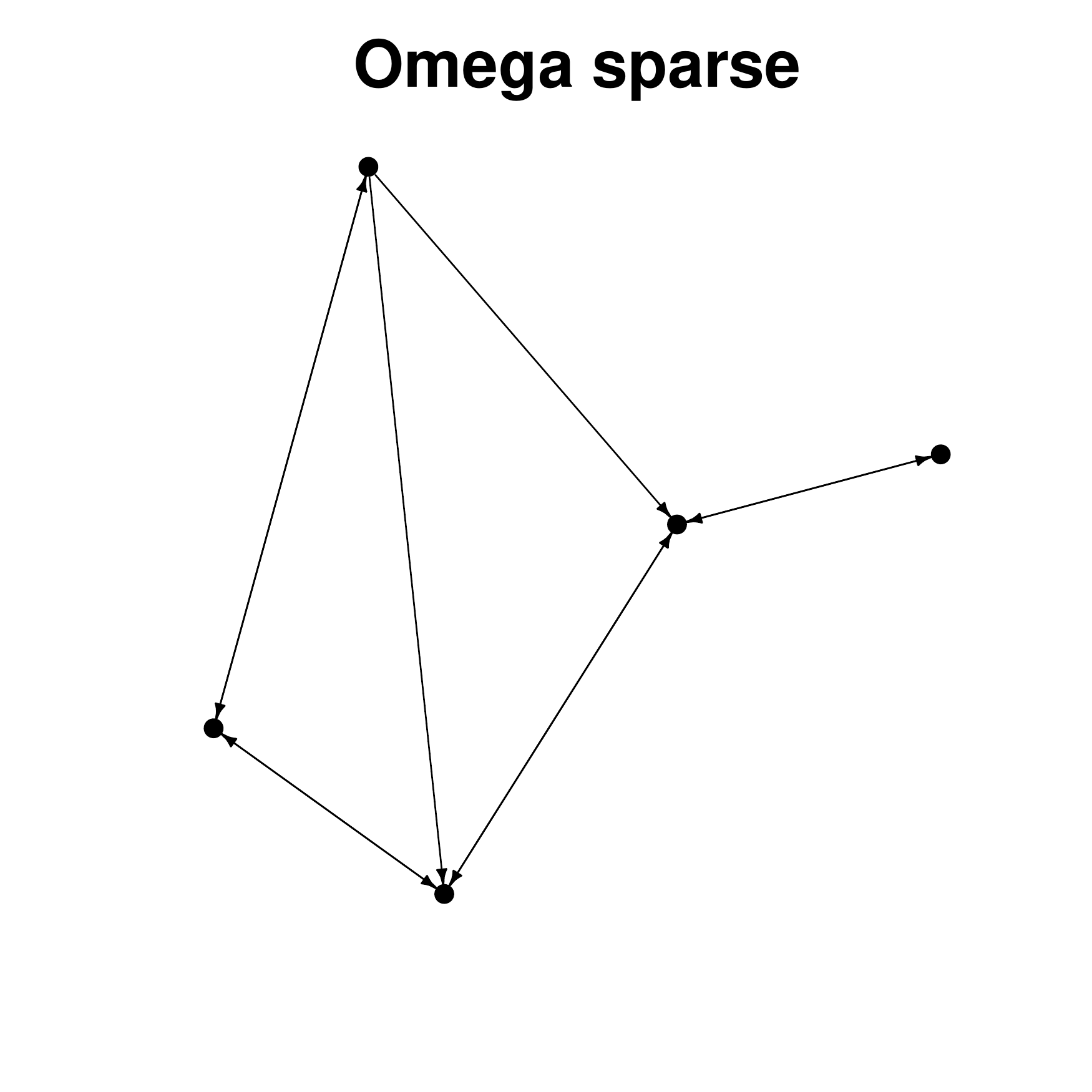}} &  \multicolumn{2}{c}{\includegraphics[width=4cm]{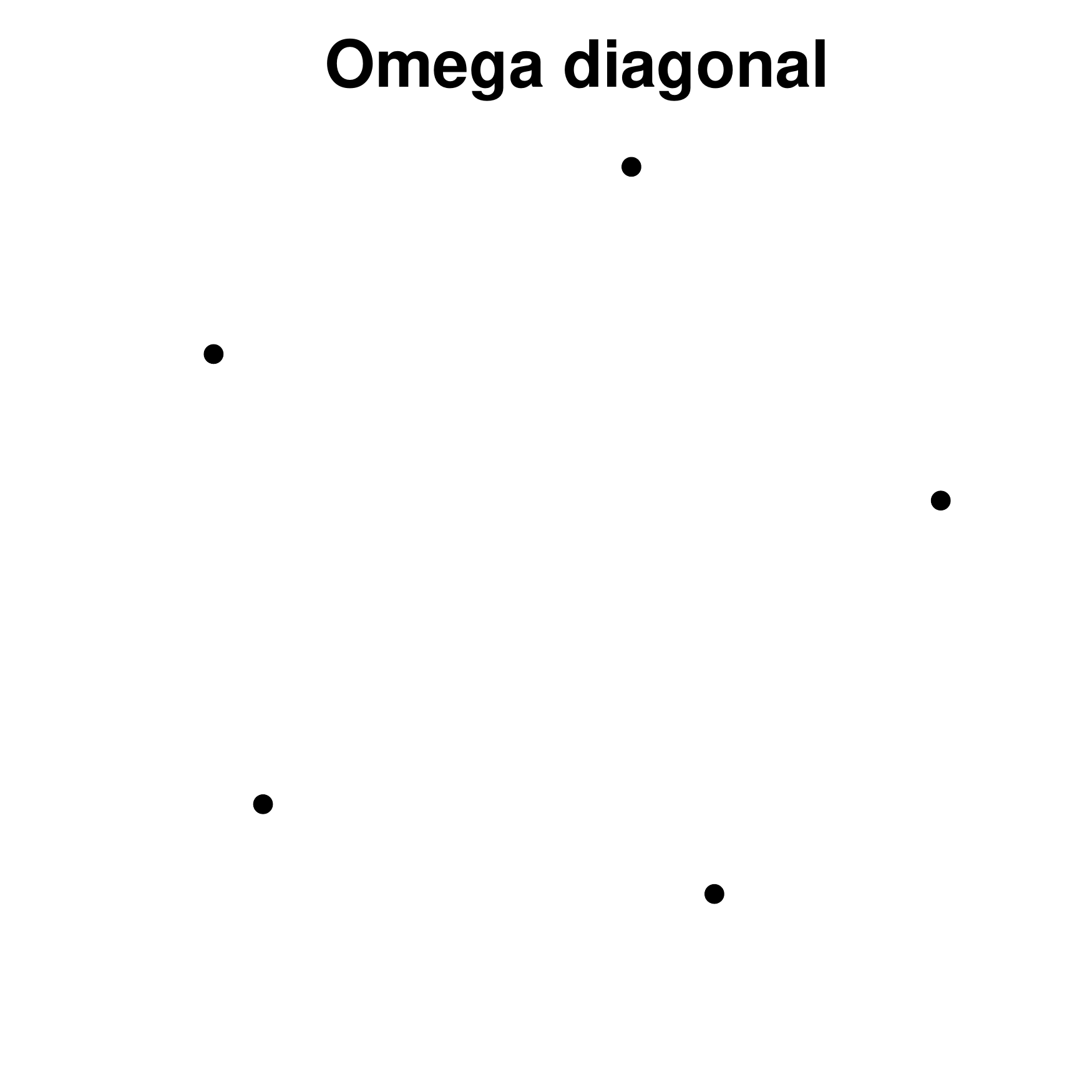}} & 
 \multicolumn{2}{c}{\includegraphics[width=4cm]{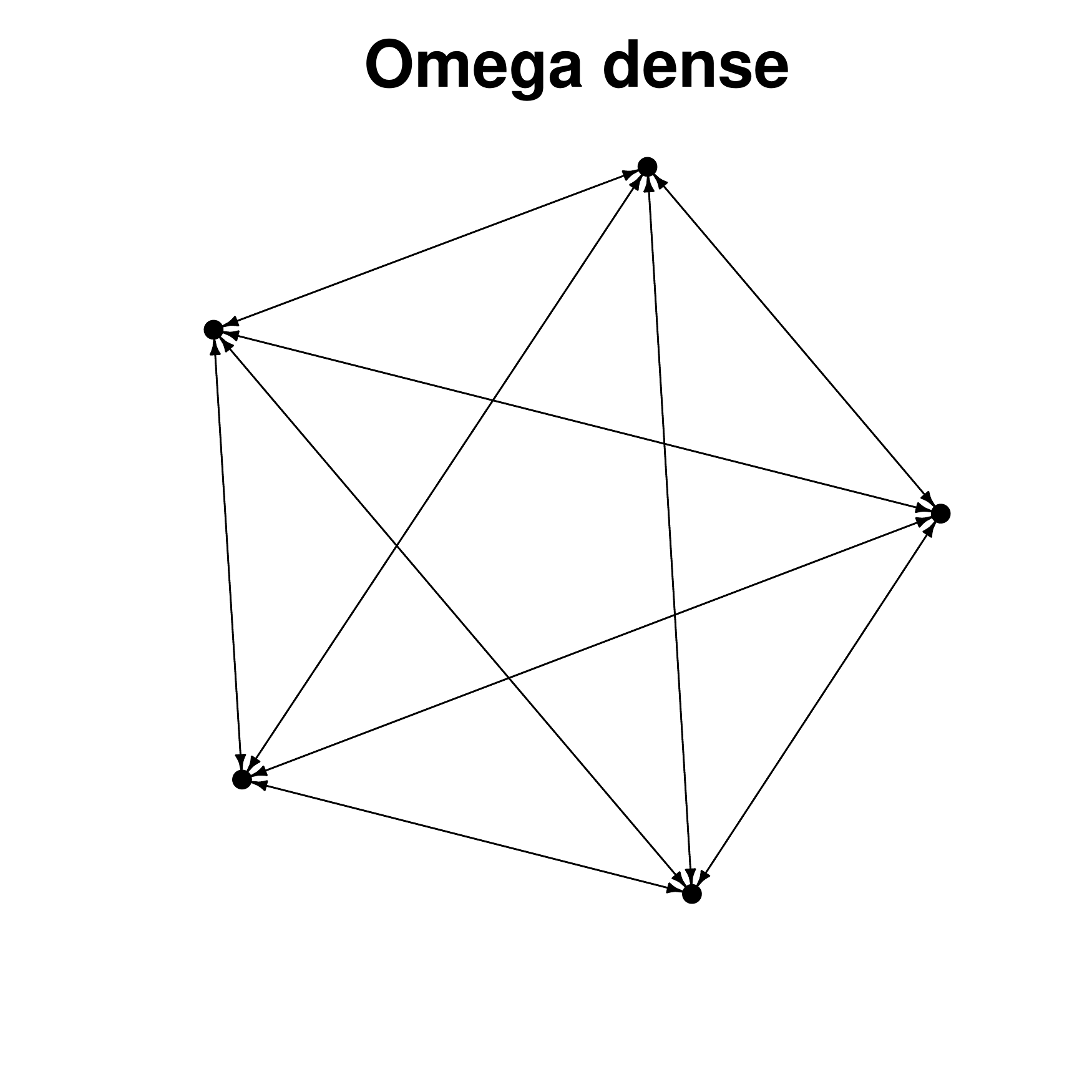}}
 \\ 
 &   &&  \multicolumn{2}{c}{$\rho=0.6$} &&&&\\  
 & Estimator && MAEE & TPR/TNR &  MAEE & TPR/TNR &   MAEE & TPR/TNR \\ \hline
$K=5$& \texttt{GroupLasso+Cov}	&&   0.251 & 1.00/0.62 & 0.253  & 1.00/0.61 & 0.244  & 1.00/0.67 \\ 
& \texttt{GroupLasso} 		&&   0.379 & 1.00/0.53 & 0.349 & 1.00/0.54 & 0.394 & 1.00/0.53 \\ 
& \texttt{Lasso+Cov} 					&&   0.285 & 0.91/0.90 & 0.286 & 0.91/0.88 & 0.282 & 0.91/0.91 \\ 
& \texttt{Lasso}				&&   0.497 & 0.96/0.67 & 0.303 & 0.91/0.86 & 0.374 & 0.91/0.85 \\ 
&&&&&&&&\\
$K=20$& \texttt{GroupLasso+Cov}  	&&   0.156 & 1.00/0.35 & 0.155  & 1.00/0.35 & 0.155 & 1.00/0.35\\ 
& \texttt{GroupLasso}     	&&   0.470 & 1.00/0.15 & 0.411  & 1.00/0.36 & 0.503& 1.00/0.15\\
& \texttt{Lasso+Cov} 					&&   0.281 & 0.96/0.69 & 0.279  & 0.96/0.69 & 0.281 & 0.96/0.69\\ 
& \texttt{Lasso} 				&&   0.547 & 0.96/0.56 & 0.436  & 0.96/0.57 & 0.589 & 0.96/0.56\\  
&&&&&&&&\\
$K=50$ & \texttt{GroupLasso+Cov}  	&&   0.196 & 1.00/0.40 & 0.196  & 1.00/0.40  & 0.196 & 1.00/0.40 \\ 
& \texttt{GroupLasso} 			&&   0.353 & 1.00/0.35 & 0.351  & 1.00/0.35 & 0.353 & 1.00/0.35\\ 
& \texttt{Lasso+Cov} 					&&   0.259 & 0.89/0.73 & 0.260 & 0.89/0.73 & 0.259  & 0.89/0.73\\     
& \texttt{Lasso} 				&&   0.526 & 0.89/0.71 & 0.525 & 0.89/0.71 & 0.529 & 0.89/0.71\\     \hline
\end{tabular}
\end{minipage} }
\end{table}

The MAEE for all simulation designs are reported in Table \ref{CatTable}.
In line with Figure \ref{MAEECATrho},  \verb|GroupLasso+Cov| provides a considerable improvement in MAEE over  \verb|GroupLasso| when the error terms are correlated, see ``Omega sparse", with $\rho=0.6$. For reasons of brevity, we only report the results for $\rho=0.6$. The estimation accuracy improves by more than 30\%. 
The improvement of \verb|GroupLasso+Cov| over \verb|GroupLasso| becomes even larger when the number of categorical regressors $K$ increases.
A paired $t$-test confirms that this improvement is significant (all $p-$values $<0.01$). %The \verb|GroupLasso+Cov| also significantly outperforms both lasso estimators.

When $\boldsymbol\Omega$ is diagonal or dense, \verb|GroupLasso+Cov| also attains the best estimation accuracy. Even though  $\boldsymbol{\Omega}$ is not sparse in the latter setting, and our proposed estimator provides a sparse estimate of  $\boldsymbol{\Omega}$, it still provides a considerable improvement over the \verb|GroupLasso| by exploiting the correlated error term structure.
Furthermore, the \verb|GroupLasso+Cov| also significantly outperforms both lasso estimators.

Table \ref{CatTable} also contains the results on the True Positive Rate and True Negative Rate. 
The \verb|GroupLasso+Cov| performs very similar to the \verb|GroupLasso|. Accounting for the error correlation mainly affects the estimation accuracy, but only to a lesser extent the sparsity recognition performance. A similar observation is made by \cite{Rothman10}.
Furthermore, the group lasso estimators attain, overall, a higher true positive rate than the lasso estimators. %The latter attain, overall, a higher true negative rate than the former. 

\bigskip

\begin{figure}[t]
\centering
\includegraphics[width=12cm]{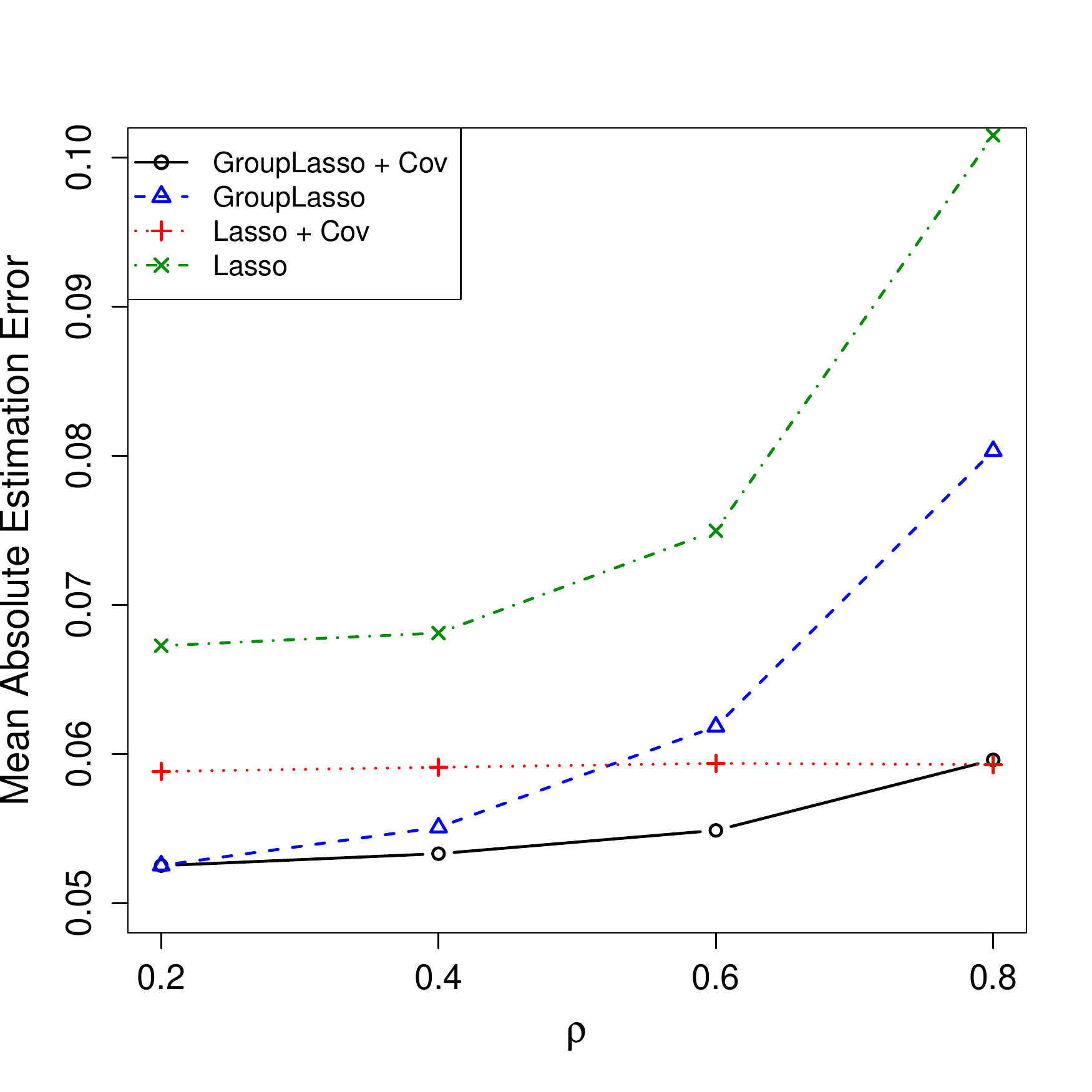}
\caption{\small VAR(2) model of dimension $q=5$ and $T=50$: Mean Absolute Estimation Error versus the correlation $\rho$, for the four considered estimators. \label{MAEETSrho}}
\end{figure}

\noindent
{\bf Time series.}
First consider the settings with a sparse inverse error covariance structure. The MAEE for the VAR(2) model of dimension $q=5$ is displayed in Figure \ref{MAEETSrho} for different values of $\rho$. We find that 
(i) the improvement of \verb|GroupLasso+Cov| over \verb|GroupLasso| is remarkable when the error terms are highly correlated,  
%the \verb|GroupLasso+Cov| attains a better estimation accuracy than the \verb|GroupLasso|. The improvement in estimation accuracy is remarkable when the error correlation $\rho$ is large.
(ii) \verb|GroupLasso+Cov| and \verb|GroupLasso| perform similarly when the error terms are hardly correlated,  
(iii) the group lasso estimators perform, overall, better than the corresponding lasso estimators.

The MAEE for all simulation designs are reported in Table \ref{TSTable}. %For reasons of brevity, we only report the results for $\rho=0.6$ under the sparse inverse error covariance structure.
For correlated errors (cfr. ``Omega sparse" and ``Omega dense"), the \verb|GroupLasso+Cov| performs best and attains, in general, a considerably lower MAEE than the \verb|GroupLasso|.
For uncorrelated errors (``Omega diagonal"), the differences in estimation accuracy between \verb|GroupLasso+Cov| and \verb|GroupLasso| are less outspoken. Importantly, there is no loss in using the former compared to the latter. By sparsely estimating $\boldsymbol{\Omega}$, the absence of error correlation is accounted for. 

\begin{table}[t]
\small
\caption{\small VAR(2) of dimension $q \in \{ 5, 20, 50\}$ and $T=50$: Mean Absolute Estimation Error, True Positive and True Negative Rate.  \label{TSTable}}
\resizebox{0.85\textwidth}{!}{\begin{minipage}{\textwidth}
\centering
\begin{tabular}{lllcccccc} \hline
 &&& \multicolumn{2}{c}{\includegraphics[width=4cm]{Omegasparse}} &  \multicolumn{2}{c}{\includegraphics[width=4cm]{Omegadiag}} & 
 \multicolumn{2}{c}{\includegraphics[width=4cm]{Omegadense}}
 \\ 
 &   &&  \multicolumn{2}{c}{$\rho=0.6$} &&&&\\  
 & Estimator && MAEE & TPR/TNR &  MAEE & TPR/TNR &   MAEE & TPR/TNR \\ \hline
$q=5$& \texttt{GroupLasso+Cov} 		&&   0.055 & 0.86/0.77 & 0.053 & 0.87/0.89  & 0.058  & 0.85/0.66 \\ 
& \texttt{GroupLasso}		&&   0.062 & 0.87/0.75 & 0.051 & 0.87/0.89 & 0.072 & 0.86/0.64\\ 
& \texttt{Lasso+Cov}					&&   0.059 & 0.79/0.62 & 0.059 & 0.80/0.69 & 0.059 & 0.78/0.55\\ 
& \texttt{Lasso} 				&&   0.075 & 0.54/0.92 & 0.068 & 0.49/0.97 & 0.090 & 0.55/0.86\\ 
&&&&&&&&\\
$q=20$& \texttt{GroupLasso+Cov}  		&&  0.015  & 0.86/0.64 & 0.015 & 0.83/0.76 & 0.017 & 0.89/0.49\\ 
& \texttt{GroupLasso}     				&&  0.024  & 0.87/0.54 & 0.018  & 0.84/0.71 & 0.044 & 0.90/0.36\\
& \texttt{Lasso+Cov} 					&&  0.015  & 0.78/0.51 & 0.015  & 0.76/0.61 & 0.016 & 0.80/0.42\\ 
& \texttt{Lasso} 						&&  0.028  & 0.52/0.84 & 0.019  & 0.47/0.91 & 0.069 & 0.58/0.71\\  
&&&&&&&&\\
$q=50$& \texttt{GroupLasso+Cov}  		&&  0.006  & 0.68/0.92 & 0.006 & 0.67/0.95   &  0.008 &  0.73/0.80\\ 
& \texttt{GroupLasso} 					&&  0.007  & 0.68/0.92 & 0.006 & 0.67/0.95 & 0.019 & 0.73/0.80\\ 
& \texttt{Lasso+Cov} 					&&  0.006  & 0.61/0.36 & 0.006  & 0.62/0.84 & 0.007 & 0.62/0.76\\     
& \texttt{Lasso} 						&&  0.007  & 0.83/0.98 & 0.006 & 0.34/0.98 & 0.027& 0.43/0.92\\     \hline
\end{tabular}
\end{minipage} }
\end{table}

%As for the categorical data, the group lasso estimators, overall, outperform the lasso estimators in terms of estimation accuracy. 
Differences in the sparsity recognition between the estimators are less outspoken. While the estimators perform more similarly in terms of sparsity recognition, the considerable improvement in estimation accuracy attained by the \verb|GroupLasso+Cov| gives it a clear advantage over the other estimators.

\section{Application \label{Application}}
We consider a data set of 30 mammary gland gene expression variables of mice \citep{Abegaz13}. Data are available for 18 time points, so we estimate a VAR(2) model of dimension $q=30$, with $T=18$. Since three samples are available, we estimate the VAR model three times. 

We make an out-of-sample forecast comparison between  \verb|GroupLasso+Cov|, \verb|GroupLasso|, \verb|Lasso+Cov|, and \verb|Lasso|. We use an expanding window approach. For $t=13,\ldots,T-1$, we estimate the VAR(2) model using time points one until $t$ and compute the one-step-ahead forecast. We compare the performance of the different estimators using the Mean Absolute Forecast Error
\begin{equation}
\text{MAFE} = \dfrac{1}{5} \sum_{t=13}^{T-1} \dfrac{1}{q}\sum_{i=1}^{q} |y^{(i)}_{t+1} - \widehat{y}^{(i)}_{t+1}|,
\end{equation}
where $\widehat{y}^{(i)}_{t+1}$ is the estimate of the $i^{th}$ response at time $t+1$.
We repeat this exercise three times, once for each replicate sample. Results are given in Table \ref{Forecast}.

The \verb|GroupLasso+Cov| attains the best forecast performance. It is closely followed by the \verb|Lasso+Cov|. An important gain in prediction accuracy is obtained by accounting for the correlation structure of the error terms: the MAFE of the \verb|GroupLasso+Cov|  is, on average, 45\% lower than the MAFE of  the \verb|GroupLasso|. Furthermore, we see from Table \ref{Forecast} that the group lasso estimators perform better than the corresponding lasso estimators.

\begin{table}
\small
\caption{\small Mean Absolute Forecast Error for the four considered estimators (rows) and three samples (column). The average MAFE, averaged over the three samples, is provided in the last column. \label{Forecast}}
\centering
\begin{tabular}{lllcccc} \hline
Estimator &&& Sample 1 & Sample 2 & Sample 3 & Average \\  \hline
\texttt{GroupLasso+Cov} 		&&& 0.81 & 0.80 & 0.80 & 0.80\\
\texttt{GroupLasso} 	&&& 1.23 & 1.38 & 1.72 & 1.44\\
\texttt{Lasso+Cov} 			&&& 0.83 & 0.81 & 0.97 & 0.87 \\
\texttt{Lasso} 		&&& 1.51 & 1.85 & 2.37 & 1.91\\  \hline
\end{tabular}
\end{table}

\begin{figure}[t]
\centering
\includegraphics[width=10cm]{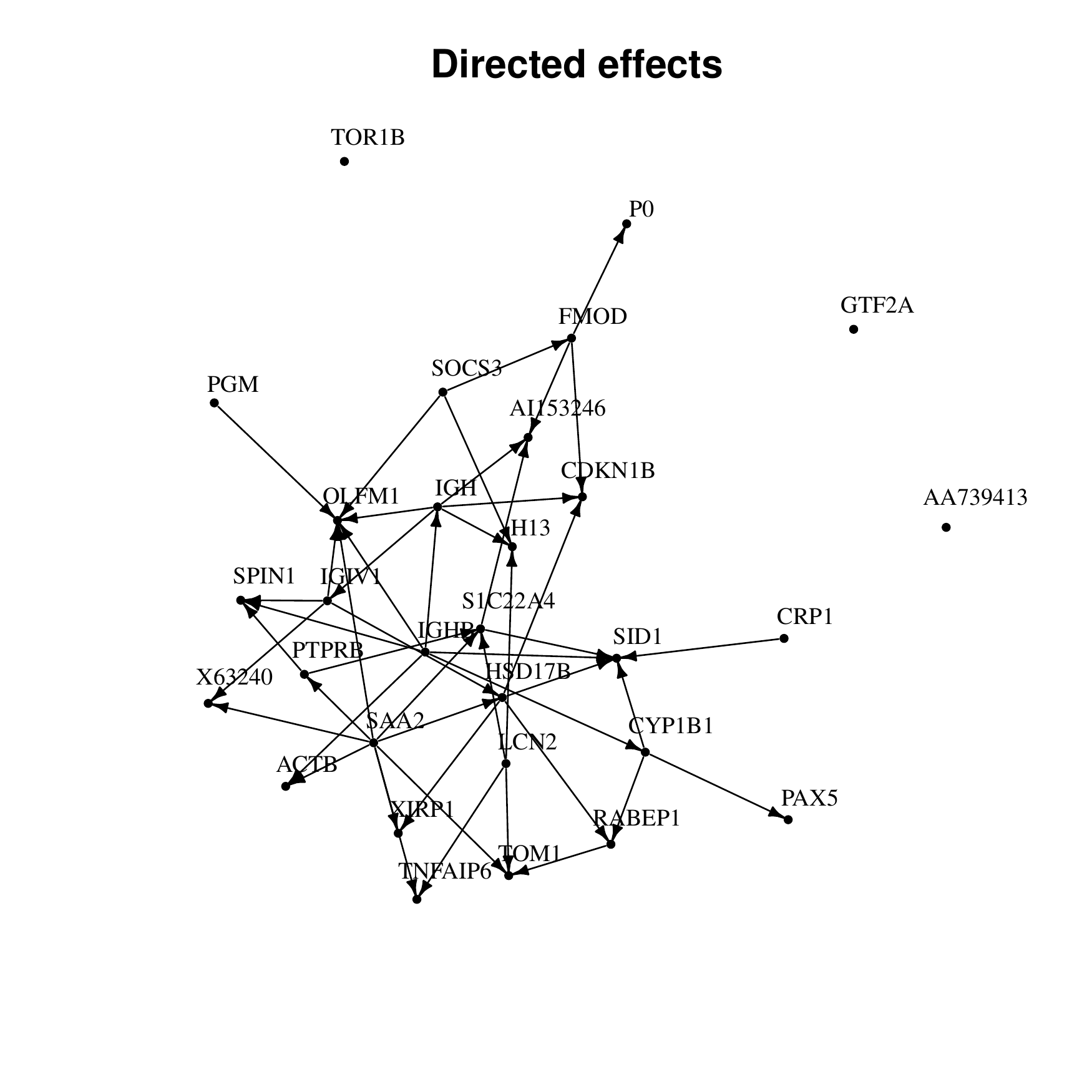} 
\vspace{-1cm}
\caption{\small Directed effects: a directed edge is drawn from one gene to another if  the \texttt{GroupLasso+Cov} estimator indicates, by giving a non-zero regression estimate, that the former influences the latter. \label{GeneBeta}}
\end{figure}

We study the interaction between the genes that trigger transitions to the mammary gland's main development stages.
%The \verb|GroupLasso+Cov| provides a simultaneous estimate of the regression coefficients and the inverse error covariance matrix. This way, we can study 
%(i) the directed effects in the gene network, inferred from the estimated regression coefficients $\bf{\widehat{B}}$ and 
%(ii) the undirected effects between the genes, inferred from the estimated inverse error covariance matrix $\boldsymbol{\widehat{\Omega}}$.
Figure  \ref{GeneBeta} represents the ``directed, lagged effects" \citep{Abegaz13} inferred from $\bf{\widehat{B}}$. We discuss the results obtained from the first sample. Results for the other two samples are similar and  available from the authors upon request. The nodes in the network are the different genes. A directed edge from gene A to gene B is drawn if the \verb|GroupLasso+Cov| indicates, by giving a non-zero estimate, that gene A has a  lagged effect on gene B.
The solution is very sparse: 850 out of the possible $900=30^2$ effects are estimated as zero. Some genes such as $\verb+GTF2A+$ and $\verb+TOR1B+$, neither influence any other genes, nor are influenced by other genes. Other genes, such as $\verb+HSD17B+$ and $\verb+SAA2+$ are important hubs in the gene regulatory network. Previous research (\citealp{Abegaz13} and references therein)  found  these genes to play a central role in the mammary gland's development stages. 

\begin{figure}[t]
\centering
\includegraphics[width=9cm]{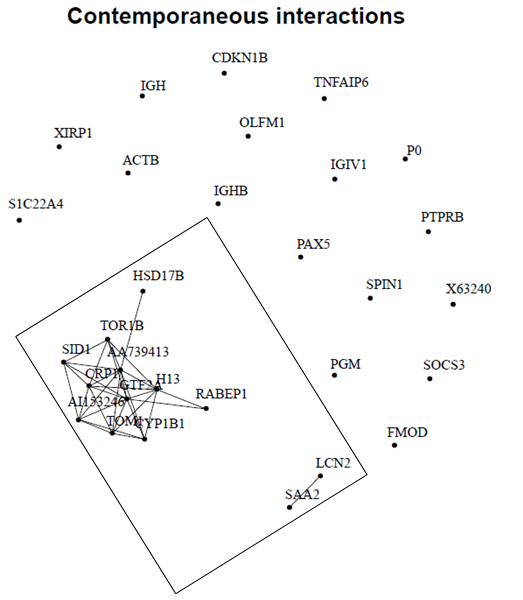}
%\vspace{-1cm}
\caption{\small Contemporaneous interactions: an undirected edge is drawn between two genes if the \texttt{GroupLasso+Cov} estimator indicates, by giving a non-zero estimate in $\boldsymbol{\Omega}$, that the innovations are partially correlated. Contemporaneous interactions are observed for only a subset of 13 genes, as indicated by the rectangle.  \label{GeneOmega}}
\end{figure}

Figure \ref{GeneOmega} represents the ``contemporaneous interactions"  \citep{Abegaz13} inferred from $\boldsymbol{\widehat{\Omega}}$. Again, the genes are the different nodes in the network. The elements of $\boldsymbol{\widehat{\Omega}}$ have a natural interpretation as partial correlations between the innovations (or error components) of the $q$ equations in the VAR model. An edge is drawn between gene A and gene B if the corresponding element in the inverse error covariance matrix is estimated as non-zero.  This means that the innovations of genes A and B are contemporaneously partially correlated: conditional on all other innovations, a shock in the innovation of gene A will lead to an instantaneous shock in the innovation of gene B, and vice versa. As can be seen from Figure \ref{GeneOmega}, contemporaneous interactions are observed only between a subset of 13 gene innovations, indicated by the rectangle.  %The right panel of Figure \ref{GeneOmega} zooms in on those 13 genes.  
An important advantage of the sparse estimator is that the main interactions in the large gene regulatory network are highlighted. Out of the possible 435 interactions, only  32 are estimated as non-zero. As such, the researcher can concentrate on these results to further deepen our knowledge into the interactions at play in the development stages of the mammary gland.

%\section{Conclusion \label{Conclusion}}
%In regression settings where the predictors are grouped, one typically wants to select important predictor groups rather than individual predictors. The regression model with categorical variables or the time series model with lags are typical examples. This paper focuses on the multivariate linear regression model where the predictors exhibit a natural grouping. We provide a group lasso estimator for the multivariate regression model that explicitly accounts for the multivariate nature of the model by exploiting the correlation between the error terms.
%We show in several simulation studies and a genetics data application that the proposed method outperforms the group lasso that does not account for the error correlation.

\bigskip

\noindent
{\bf Acknowledgments.} The authors gratefully acknowledge financial support from the FWO (Research Foundation Flanders, contract number 11N9913N).

\bibliographystyle{asa}
\bibliography{bibfile}
     
\end{document}